\documentclass[conference]{IEEEtran}
\IEEEoverridecommandlockouts
\usepackage{cite}
\usepackage{amsmath,amssymb,amsfonts}
\usepackage{algorithmic}
\usepackage{graphicx}
\usepackage{textcomp}
\usepackage{xcolor}
\usepackage{caption}

\usepackage{comment}

\usepackage{url}
\usepackage{courier}
\usepackage{booktabs}
\usepackage{listings}
\usepackage{wrapfig}
\usepackage{blindtext}% for example text here only
\usepackage{adjustbox}
\usepackage{balance}
\usepackage{float}
\usepackage{hyperref}

\def\BibTeX{{\rm B\kern-.05em{\sc i\kern-.025em b}\kern-.08em
    T\kern-.1667em\lower.7ex\hbox{E}\kern-.125emX}}
\begin{document}

\title{Towards Provenance-Aware Earth Observation Workflows: the openEO Case Study}

% redefine command to show numbers instead of symbols
\DeclareRobustCommand{\IEEEauthorrefmark}[1]{\smash{\textsuperscript{\footnotesize #1}}}

\author{\IEEEauthorblockN{
H. Omidi*\IEEEauthorrefmark{1},  
L. Sacco*\IEEEauthorrefmark{1}, 
V.
Hutter\IEEEauthorrefmark{2}, 
G. Irsiegler\IEEEauthorrefmark{2}, 
M. Claus\IEEEauthorrefmark{3},
M. Schobben\IEEEauthorrefmark{4},\\
A. Jacob\IEEEauthorrefmark{3}, 
M. Schramm\IEEEauthorrefmark{4}, 
S. Fiore\IEEEauthorrefmark{1}
}
\IEEEauthorblockA{\IEEEauthorrefmark{1}University of Trento, Trento, Italy \\
\IEEEauthorrefmark{2}Earth Observation Data Centre for Water Resources Monitoring GmbH, Vienna, Austria \\
\IEEEauthorrefmark{3}Eurac Research, Bolzano, Italy  \\
\IEEEauthorrefmark{4}Vienna University of Technology, Vienna, Austria}

\thanks{* The two authors equally contributed to this work.}}

\maketitle

\begin{abstract}

Capturing the history of operations and activities during a computational workflow is significantly important for Earth Observation (EO).
The data provenance helps to collect the metadata that records the lineage of data products, providing information about how data are generated, transferred, manipulated, by whom all these operations are performed and through which processes, parameters, and datasets. 
This paper presents an approach to improve those aspects, by integrating the data provenance library yProv4WFs within openEO, a platform to let users connect to Earth Observation cloud back-ends in a simple and unified way. In addition, it is demonstrated how the integration of data provenance concepts across EO processing chains enables researchers and stakeholders to better understand the flow, the dependencies, and the transformations involved in analytical workflows. 
\end{abstract}

\begin{IEEEkeywords}
Provenance, Earth Observation, openEO, WfMS
\end{IEEEkeywords}

\section{Introduction}
Provenance, as a general concept, refers to the history of ownership of a valued object, and more specifically, data provenance refers to the process of tracking and capturing the origins and movement of data in the way we track the whole lifetime of data throughout a process \cite{b9}.
In other words, data provenance could be considered as the metadata that records the lineage of data products.
The tracking of provenance at the computational workflow level guarantees the reproducibility and reliability of research processes in eScience \cite{b4}. 
This plays a critical role in Earth Observation (EO), as data is often processed through complex, multistage workflows that involve large-scale datasets, different data sources, and distributed processing tools.

The concept of a workflow typically refers to a high-level overview of sequentially related procedures, encompassing all the steps involved. In the context of any application or activity pipeline, a workflow represents the complete sequence of operations throughout its life-cycle. Based on that, it is possible to provide valuable information about the whole procedure by capturing data and metadata about each activity and step.
When talking about EO applications, tracking provenance can become a complicated task. The cost and complex nature of EO experiments make them difficult to monitor, consequently hindering transparency, reliability, and repeatability. A practical solution to improve these aspects is tracking provenance at the workflow level, which means capturing not only what (inputs and outputs) is produced during the workflow execution, but also how, where, when, by whom, and why each computational step has been carried out.
As noted by Tan et al. \cite{b1}, this detailed form of provenance is essential for tracking changes in EO training pipelines, especially when workflows are reused or updated over time. In fact, tracked provenance makes research activities easier for scientists working on the same computational procedures, consistently reducing the efforts needed to reproduce scientific results.
With the rise of federated EO processing systems, such as those enabled by openEO \cite{b10}, tracking the provenance at the workflow level has become even more important. Federated environments provide the processing across multiple back-ends and make it challenging to trace how each component contributes to the final output. The provenance of the workflow level is fundamental to ensure reproducibility in these environments, as it allows users to reconstruct complete end-to-end processing chains across diverse infrastructures \cite{b2}.

openEO has developed an API based on different technologies for big Earth Observation in a simple and unified way, handling large amounts of data in the EO applications, giving an easy access to EO data and the necessary compute resources and functionality needed to work with such data. Specifically, it provides different back-ends to its users, also offering APIs developed under various programming languages as Python, R, and JavaScript.
Moreover, it lets researchers focus on analytical tasks rather than data logistics. It makes Earth Observation access and processing standardized across different cloud platforms, although it also includes a version that can run on the local machine.
%\cite{b10}.

%---------------yProv4OpenEO-------------------
In this work, we introduce  yProv4WFs \cite{b4} (a library which provides provenance tracking across various Workflow Management Systems (WfMSs)) and in particular its openEO extension, designed to address provenance tracking within the openEO framework.
yProv4WFs helps openEO obtain provenance records for each executed workflow, extracting runtime metrics from all stages of an openEO use case or scenario and information about tasks, data, and the relationship between them.

%--------- structure of the paper ---------
The paper is organized as follows: Section II reviews the related work while Section III introduces how to track provenance in the openEO platform. Section IV presents the yProv4WFs architecture and its components whereas Section V discusses the application of the proposed yProv4WFs extension for openEO to a real use case. Finally Section VI draws the conclusions of the paper and provides future directions.

\section{Related work}
The tracking of provenance is a critical component in ensuring reproducibility, transparency and reliability, particularly in the context of long-running and complex computational processes \cite{b25}.
The demand for open geospatial science, including both open data and open-source software, is rapidly increasing \cite{b15}. In parallel with the expansion of open geospatial platforms and the growing use of complex computational pipelines, various tools and standards have emerged to efficiently capture and represent provenance data while guaranteeing interoperability.

Scientific workflows typically consist of a series of transformations, data manipulations, and computations that span multiple data sources, software components, and computing environments.  Therefore, understanding the provenance, which includes the origin, evolution, and dependencies of data products, is critical for traceability and reusability \cite{b16}.  

In the field of EO research, reproducibility is frequently compromised due to the absence of standardized provenance, especially when combining local and cloud-based infrastructure \cite{b17}.
Ensuring the high quality and reproducibility of scientific data requires not only robust data processing infrastructures but also comprehensive provenance tracking throughout the data life-cycle. In \cite{b26}, a work on the  Atmospheric Radiation Measurement (ARM) program shows the critical importance of transparent quality control in atmospheric data streams. Recent efforts to improve scientific reproducibility and transparency have led to the development of structured, open workflows. For example, Arnal et al. \cite{b27} introduced FROSTBYTE, a reproducible and modular forecasting pipeline implemented through Jupyter Notebooks and publicly shared on GitHub. Other studies have also emphasized reproducibility; for instance, Cuevas-Vicenttín et al. \cite{b28} describe how scientific workflows can enhance both reproducibility and interoperability by automating computations and capturing provenance data.
Although several provenance models offer robust support for retrospective provenance, they often lack sufficient expressiveness to capture prospective provenance in workflows involving control flow logic \cite{b29}. This can reduce workflow reproducibility, particularly in EO pipelines with conditional operations.

Despite the availability of large-scale EO datasets and high-level APIs, such as openEO \cite{b2}, workflow provenance is not automatically captured during processing. Provenance standards such as W3C PROV\footnote{https://www.w3.org/TR/prov-overview/} offer an abstract data model to capture information about metadata that can be collected during the execution of workflows, but often require extensions to be customized with a specific implementation to be useful in complex workflows\cite{b19}.

The provenance interoperability of a workflow management system across diverse workflows is also a challenge. Although many provenance models aim for standardization, they often have semantic mismatches and system-specific implementations. Furthermore, complex provenance models are often challenging for non-expert users to interpret or reuse effectively.
Some contributions are identified in RO-Crate \cite{b19} and CWLProv \cite{b21}, which both provide tracked provenance for workflows using structured metadata containers and JSON-LD-based\cite{sota8} ontologies. RO-Crate, for example, separates prospective and retrospective provenance and packages workflow artifacts for reusability. Although Ro-Crate is using some terms such as \textit{HowToStep} and \textit{FormalParameter} that limit the clarity in representing workflow dependencies, especially for users unfamiliar with semantic models. 

Even though provenance tracking at the workflow level can significantly support reproducibility, it may not be sufficient in cloud-based environments. According to Hasham, Munir, and McClatchey \cite{b20}, ensuring reproducibility in cloud-based EO platforms is particularly challenging, as users often lack clear and comprehensive access to the underlying computational environment. Therefore, provenance systems designed for the cloud must capture both workflow-level and infrastructure.

In this context, we introduce a yProv4WFs extension for the openEO platform to improve usability, reproducibility, and interoperability. It is developed in compliance with the PROV-O\footnote{https://www.w3.org/TR/prov-o/}, also improving the expressiveness and navigability of the workflow steps.

\section{Tracking provenance in the OpenEO platform}
openEO is an open API designed to unify access to cloud-based Earth Observation (EO) processing services, and it also supports local processing through compatible Python libraries. It provides a standardized interface that helps the users utilize the different EO back-ends, enabling scalable and cross-platform geospatial analytics. Through client libraries in Python, R, and JavaScript, openEO abstracts the complexity of the underlying infrastructures, allowing users to construct workflows that are portable and interoperable\cite{b3}. 

\begin{figure}[h]
\centering
\includegraphics[width=0.45\textwidth]{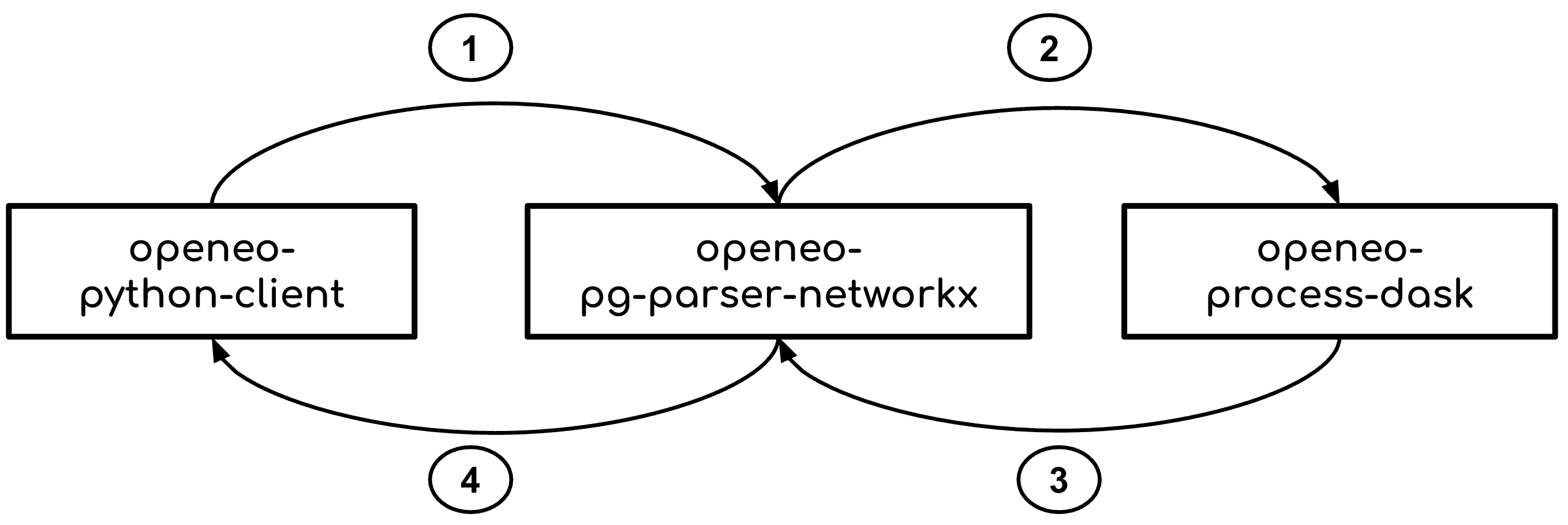}
\caption{Schematic relationship between the different packages in openEO implementation}
\label{fig:openEO_structure}
\end{figure}

The Python implementation of openEO is structured around a modular architecture consisting of three main packages: \texttt{openeo-python-client}\footnote{https://github.com/Open-EO/openeo-python-client}, \texttt{openeo-pg-parser-networkx}\footnote{https://github.com/Open-EO/openeo-pg-parser-networkx}, and \texttt{openeo-processes-dask}\footnote{https://github.com/Open-EO/openeo-processes-dask}. Each of these components plays a distinct role in the client-side definition, parsing, and execution of Earth Observation workflows. Figure \ref{fig:openEO_structure} illustrates the schematic relationship between these packages. 

\texttt{openeo-python-client} provides a high-level interface for users to connect to openEO back-ends, build workflows, and manage jobs. It abstracts communication and offers a user-friendly API for defining process graphs.
A process graph is a sequence of specific processes, each of which has its own concrete values for input parameters. Process graphs organize a pipeline of processes and automate pipeline execution\footnote{https://api.openeo.org/v/0.3.0/processgraphs/}.

\texttt{openeo-pg-parser-networkx} parses process graphs into networkx-based structures, enabling validation and traversal. It is essential for interpreting and analyzing the graph representation of workflows. Moreover, when referring to the python implementation, it produces the actual executable object, which the process graph represents.

\texttt{openeo-processes-dask} provides python implementations of openEO processes, Using the xarray and dask frameworks for scalable and efficient data processing. It is designed to be used in alongside with openeo-pg-parser-networkx, which is responsible for parsing and executing openEO process graphs.

\subsection{Tracking provenance on the Local Processing}
openEO has been developed as an API that enables users to define Earth Observation (EO) workflows using a lightweight client, while delegating the computationally intensive tasks, such as data access and processing, to the back-end systems. However, this model of remote execution and abstraction from the underlying file structures can sometimes result in workflows that fail and are challenging to debug. Each trial-and-error iteration requires establishing a connection to the back-end, waiting for the process graph to be initialized (once the necessary resources become available), and subsequently reviewing logs if the processing fails again.
To address these limitations, openEO supports local (client-side) processing. This functionality leverages the following Python libraries, integrated in the openEO Python client:
(i) \texttt{openeo-processes-dask} – An implementation of openEO processes based on Xarray and Dask, facilitating scalable, parallel processing. (ii) \texttt{openeo-pg-parser-networkx} – A parser for openEO process graphs (in JSON format) that transforms them into callable Python functions, using implementations provided by \texttt{openeo-processes-dask}.
It is important to note that these libraries are not limited to local processing. They are also integrated into certain back-end systems, such as those operated by EODC and Eurac Research. This dual usage allows users to prototype and test their openEO workflows locally, gaining insight into the expected behavior on the target back-end. As a result, users can identify and resolve issues more efficiently and with greater confidence.

\subsection{Tracking provenance on the Remote Processing}
The openEO platform allows users to outsource their processing to a list of remote backends. These backends offer a range of data sources which can be used in the creation of process graphs directly and they are generally federated so that the user can access all of them from a centralized interface.

Each backend has to expose the openEO API, defined by openEO API specification, which standardizes the formatting of requests for submitting process-graphs, managing results, starting computations and user-management, etc. Backends share a broadly overlapping structure, but the implementation details regarding compute, workflow-management or database structures may vary.

Working with remote processing closely resembles the local workflow, users create process-graphs, but instead of executing them on their own machines, the process-graphs are sent to a given backend depending on the selected data-source. 

The user can then start the processing manually and the current status (Queued, Running, Error, Finished) of the processing job will be displayed and processing logs will be made available. After the job is finished, the user will be given a list of signed URLs pointing to the resulting files which they can either visualize directly in the interface or download individually. Additionally, the Spatio Temporal Asset Catalog (STAC)\footnote{https://stacspec.org/en/about/} resources will be created for each result and exposed alongside the data (each with their own signed URL). STAC is both a standard and a collaborative community initiative aimed at enhancing access to geospatial information about our planet. Its primary objective is to simplify the process for data providers to make their datasets openly accessible and discoverable by a global audience. The provenance results will be made accessible in the same way\footnote{https://openeo.org/}.

\subsection{Workflows Execution in openEO}
It is important to distinguish between a workflow in the context of a Workflow Management System and a workflow within the openEO platform. In classic WfMSs workflows are explicit task-based execution graphs, often written in Python or a domain-specific language, which define how to execute, manage scheduling, retries, parallelism, and infrastructure. In contrast, openEO workflows are high-level declarative process graphs that describe what EO operations should be applied (e.g., filtering, aggregation, NDVI computation), without detailing how the execution is managed.

\begin{figure*}[htb!]
\centering
\includegraphics[width=0.8\textwidth]{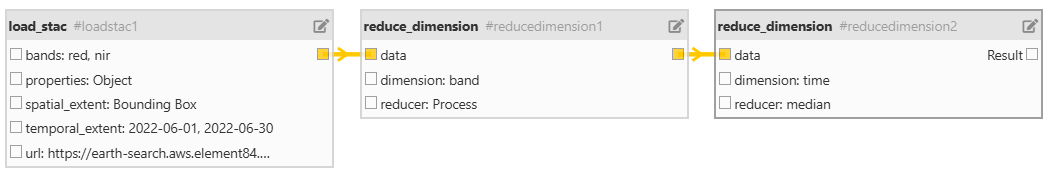}
\caption{openEO workflow with three steps}
\label{fig:three_steps_workflow}
\end{figure*}

The openEO API standard offers a unified interface for EO data processing across different back-end providers. When using the openeo-python-client\footnote{https://open-eo.github.io/openeo-python-client/} library, users build workflows by chaining operations in the form of process graphs, which are serialized to JSON and sent to a compliant back-end for execution. Users add different steps or process nodes to this workflow, each of which performs a specific task or activity.

Figure \ref{fig:three_steps_workflow} shows an example of a three steps workflow carried out by the openEO Python client module. Each box corresponds to a specific process node with its own parameters and functionality. The \texttt{load\_stac} node initializes the workflow by specifying parameters such as bands, spatial extent, temporal range, and data source URL. %Each subsequent box, like
The \texttt{reduce\_dimension} node performs a specific transformation, reducing data across a selected dimension (e.g. bands or time). The connections between boxes illustrate the data flow, with each process building on the output of the previous to form a complete modular workflow.

A typical workflow in an openEO use case includes the following steps:
\begin{itemize}
  \item Process Graph Construction: using ProcessBuilder, users create graphs that represent a series of EO operations (e.g., spatial filtering, temporal aggregation).
  \item Graph Submission: the graph is sent to a back-end. In the case the user is using the local process, the back-end would be on the local machine.
  \item Back-End Execution: the back-end interprets and executes the graph using its internal engine, which may rely on frameworks such as Dask, Airflow, or other orchestrators \cite{b3}.
  \item Result Retrieval: the output is returned as a file, visual object, or downloadable link.
\end{itemize}

\section{The yProv4WFs architecture}
\subsection{yProv}
\begin{figure}[h]
\centering
\includegraphics[width=\columnwidth]{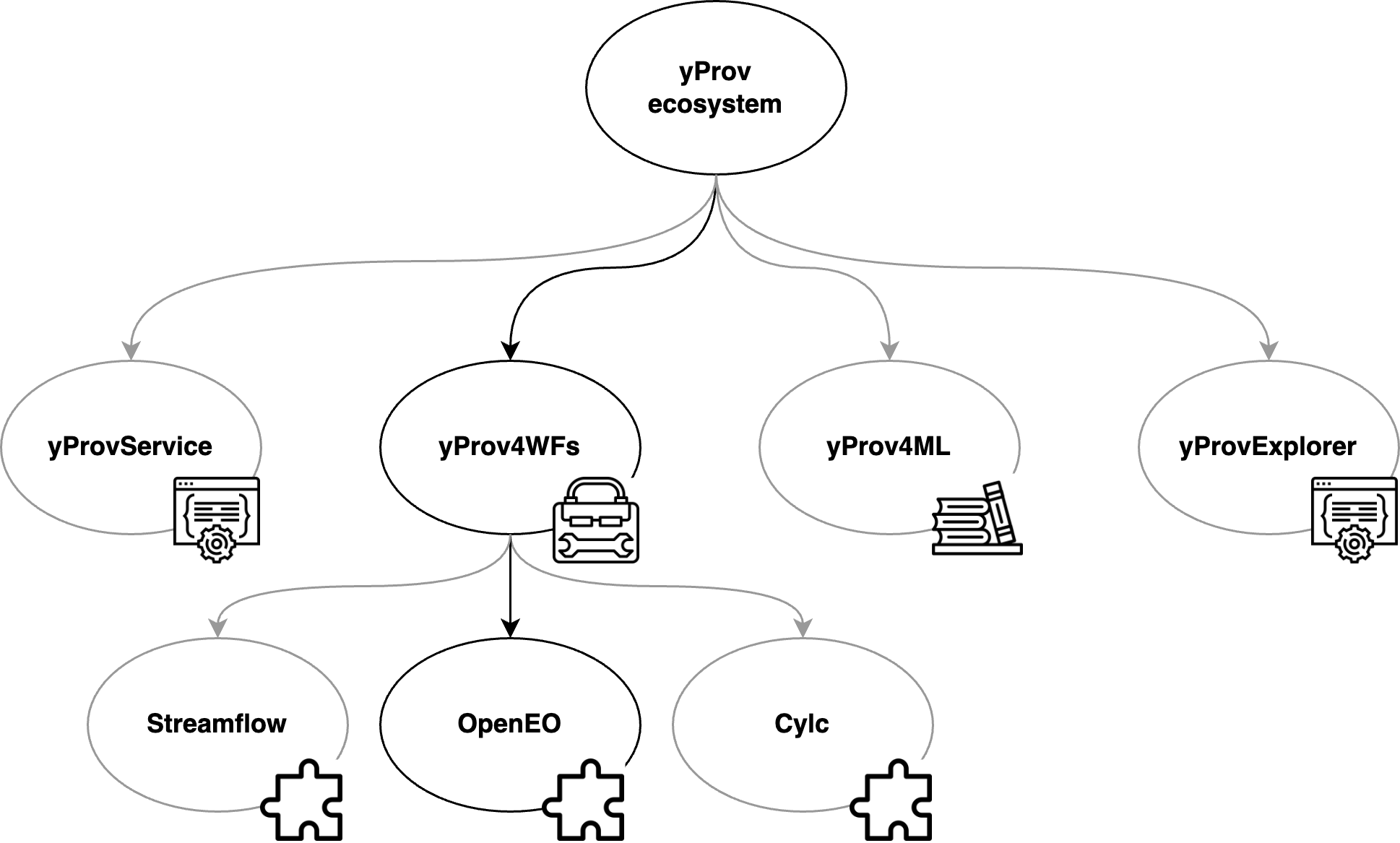}
\caption{yProv ecosystem with the different components}
\label{fig:yProv_ecosystem}
\end{figure}

yProv \cite{yProv} is an open source software ecosystem that has been developed based on the W3C PROV\footnote{https://www.w3.org/TR/prov-overview/} family of standards and contains a graph database back-end based on Neo4J. yProv is domain-agnostic and made of different components, including yProv4WFs, which has been developed for tracking provenance of workflows. Figure \ref{fig:yProv_ecosystem} shows the relationship between the different components of the yProv ecosystem.

\subsection{yProvExplorer}
The yProvExplorer\footnote{https://github.com/HPCI-Lab/yprov-explorer/} is a data science application to help scientists access and navigate provenance documents.
It is developed in Javascript and through specific libraries (i.e. React and D3.js) offers dynamic and interactive navigation of the provenance as a graph. 
All the components of provenance are distinguished by color and shape to provide a user-friendly exploring experience to the end user. It also provides other features such as navigation through history and graph statistics, and so on, which help the end user have proper tools to evaluate and compare the workflow provenance.

\subsection{yProv4WFs}
\begin{figure}[htb]
\centering
\includegraphics[width=0.45\textwidth]{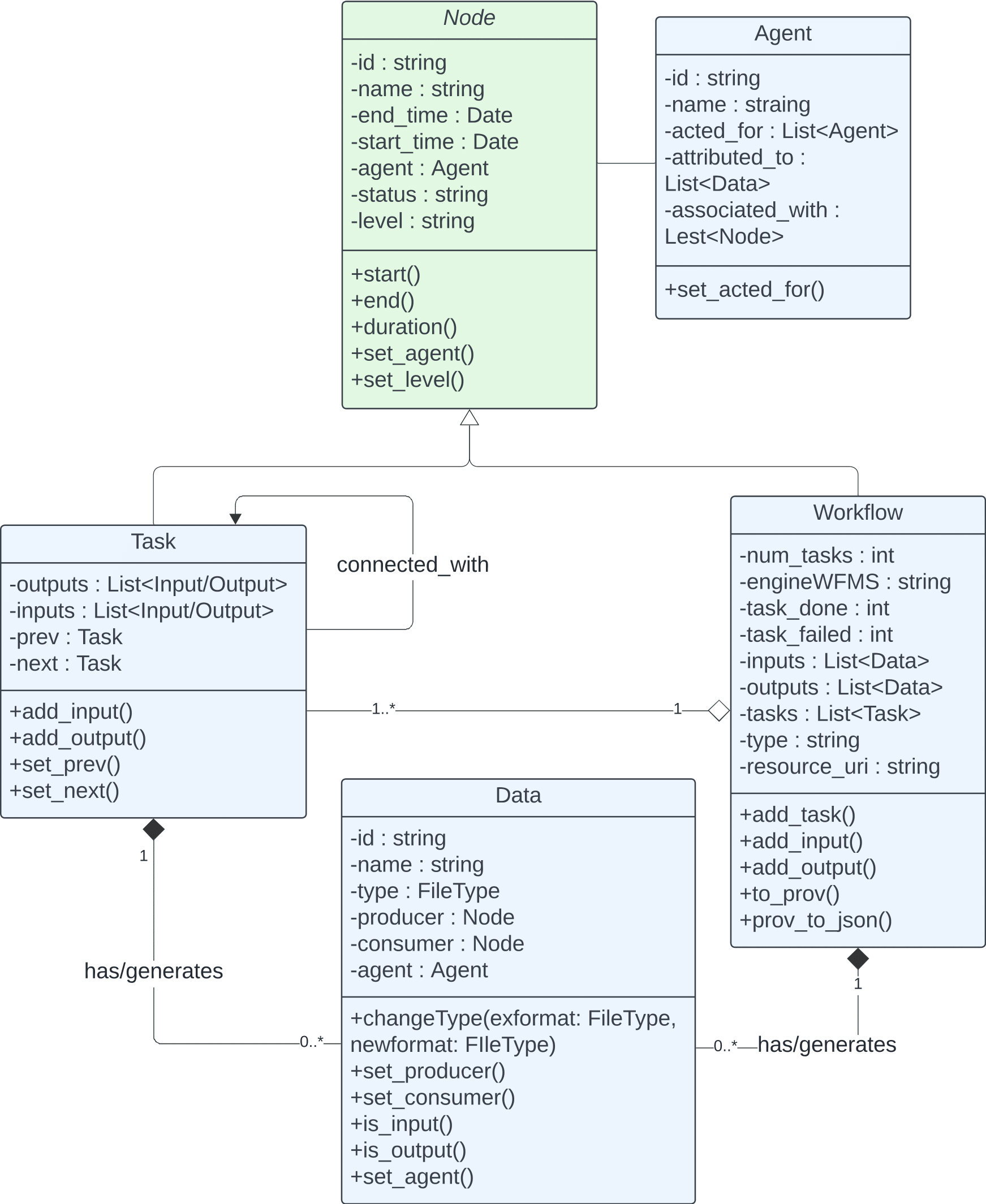}
\caption{yProv4WFs UML class diagram\cite{b4}}
\label{fig:yProv4WFs_datamodel}
\end{figure}

yProv4WFs \cite{b4} is a Python-based library to track provenance in complex scientific workflows across diverse WfMSs (see yProv4Cylc and yProv4Streamflow in  Figure \ref{fig:yProv_ecosystem}, where Cylc \cite{Cylc}  and Streamflow \cite{streamflow} are both WfMSs). It has been developed in compliance with the W3C PROV standards and designed according to the data model reported in Figure \ref{fig:yProv4WFs_datamodel}. It enables detailed and standardized recording of data lineage, task dependencies, and execution context within computational workflows. It functions as an external, third-party service, making it agnostic to specific WfMS implementations and suitable for heterogeneous, multi-system environments.

\subsection{yProv4WFs extension for openEO}

The yProv4WFs extension for openEO is designed to enable workflow-level provenance tracking in the openEO platform. It works based on the data model developed for yProv4WFs (Figure \ref{fig:yProv4WFs_datamodel}), ensuring compatibility with its provenance structure and semantics. The implementation phase of such work started on the local processing offered by the openEO API, where the execution of process graphs is simulated or executed directly in Python without involving a remote backend. This approach provided greater control and flexibility during development and debugging.

\begin{figure}[h]
\centering
\includegraphics[width=\columnwidth]{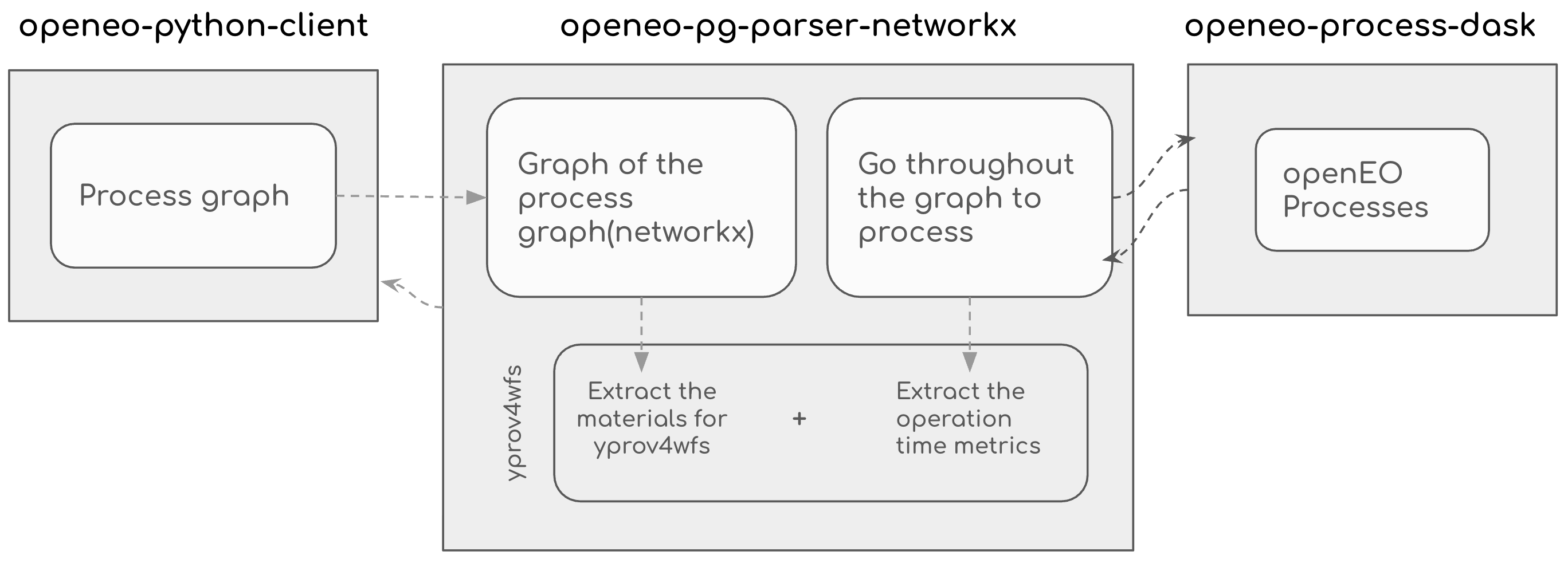}
\caption{openEO internal architecture (parser) and its integration with yProv4WFs}
\label{fig:yProv_integration}
\end{figure}

Through a deep insight of the internal structure of the openEO process graph, provenance calls have been injected into the \texttt{openeo-pg-parser-networkx} package. This package parses openEO process graphs into NetworkX DAGs, making it a natural entry point for tracking the execution flow at the node level. Figure \ref{fig:yProv_integration} shows openEO source code structure and where yProv4WFs exploits its calls to collect data and metadata information. 

The graph representation provided by the \texttt{openeo-pg-parser-networkx} package allows to systematically look over each node of the process graph, where each of them represents an EO operation or activity, and record runtime metadata such as the start time, end time, duration, and intermediate outputs. All information used to build a detailed provenance graph to reflect the actual data flow and computation within the workflow.

Following the successful and validated integration of the local processing capabilities provided by the openEO API, the focus shifted to remote execution. The same methodology was applied to enable provenance capture during the execution of jobs on remote EO infrastructures. By embedding the same node-level tracking into the back-end workflow engine, we aim to achieve consistent and transparent provenance across both local and remote executions, laying the foundation for end-to-end reproducibility and traceability in openEO workflows.

\section{Tracking provenance in the openeo-flood-mapper}
To validate the integration of yProv4WFs into the openEO platform, a set of representative EO workflows was used. Specifically, the \textit{openeo-flood-mapper}\footnote{https://github.com/interTwin-eu/openeo-flood-mapper-local} offered by the Vienna University of Technology as part of the interTwin\footnote{https://www.intertwin.eu/} project is considered for the experimental part.

\subsection{openeo-flood-mapper description}
The \textit{openeo-flood-mapper} is a Python-based module for mapping flood extent using Sentinel-1 SAR data, based on openEO commands. The module implements a Bayesian inference approach \cite{bbm22} to identify flood extents by comparing observed backscatter values with expected values over land and water surfaces. It makes use of Sentinel-1 sigma nought ($\sigma^0)$ backscatter data (level 1C) organized in a data cube, enabling time series analysis on a per-pixel basis.

Since not only water surfaces lead to low backscatter values, a contextual and probabilistic approach is required to distinguish water from non-water areas. Therefore, the expected backscatter values of non-flooded land are modeled using harmonic regression on long-term historical Sentinel-1 observations. This allows the algorithm to capture seasonal backscatter variations and estimate the expected value of a non-flooded pixel for a specific day-of-year~\cite{bbm22}. This information, as well as backscatter statistics of known water pixels, is used in a Bayesian approach for estimating the probability of a flooding. In turn, a Bayesian decision comparing posterior probabilities of flood versus non-flood allows the classification of a pixel being flooded.

Finally, the flood mapper tracks changes on the Earth’s surface over time, investigating temporal variations in Sentinel-1 SAR backscatter data to monitor surface dynamics. By examining time series at selected locations, the study aims to identify consistent trends and anomalies in surface reflectivity. This shows the importance of decision reliability by proposing the denial of decisions under conditions of high uncertainty, thus reducing the risk of misclassifications or erroneous interpretation in downstream applications.

\begin{figure*}[htb!]
    \centering  
    \includegraphics[width=\textwidth]{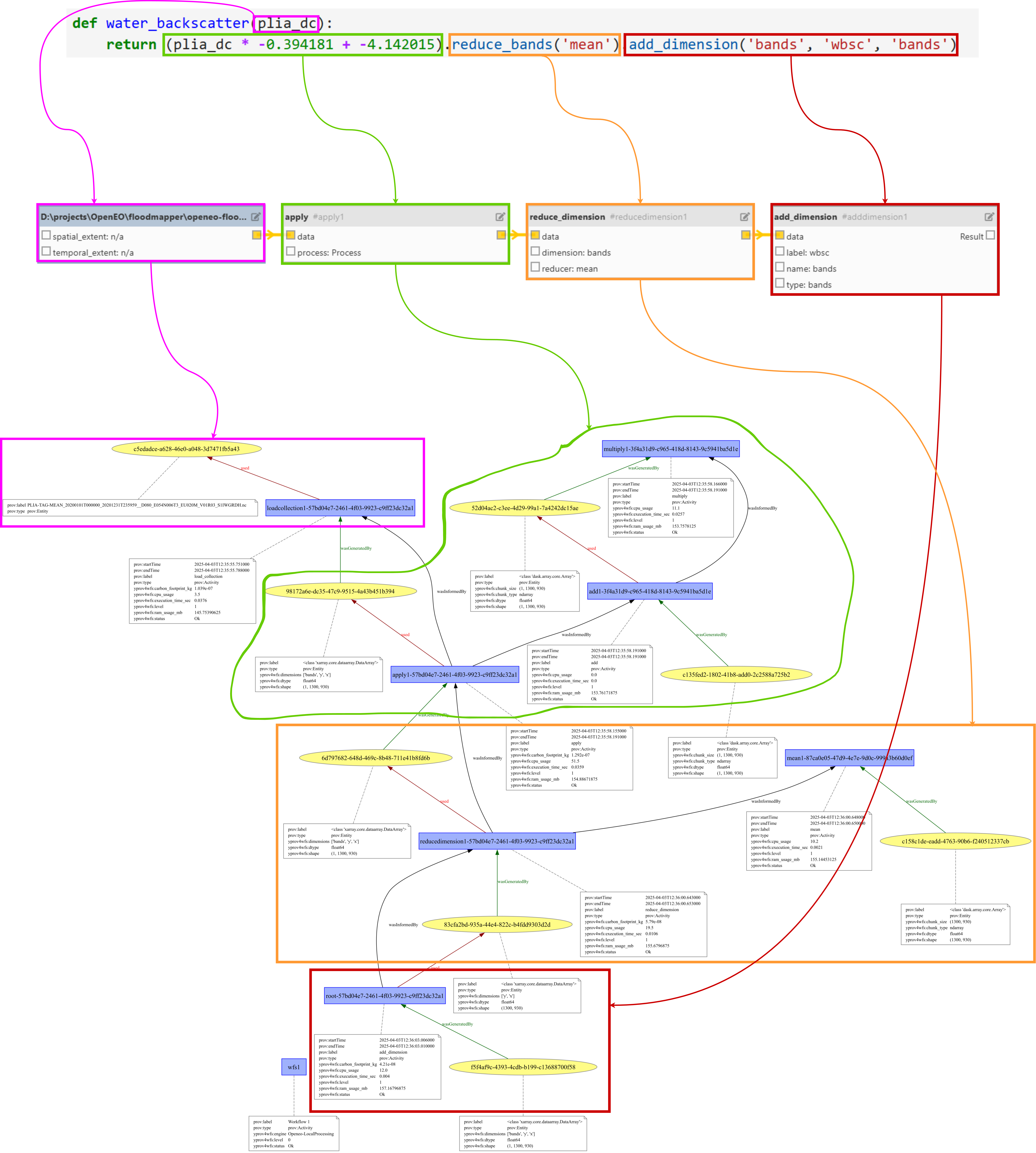}
    \caption{Representation of the \textit{water\_backscatter} Python function into an openEO process graph and in its corresponding provenance graph using yProv4WFs.}
    \label{fig:mapping_funtions}
\end{figure*}

\subsection{Provenance output on different infrastructure}

\begin{figure*}[htb!]
    \centering
    \includegraphics[width=1.2\textwidth,angle=90]{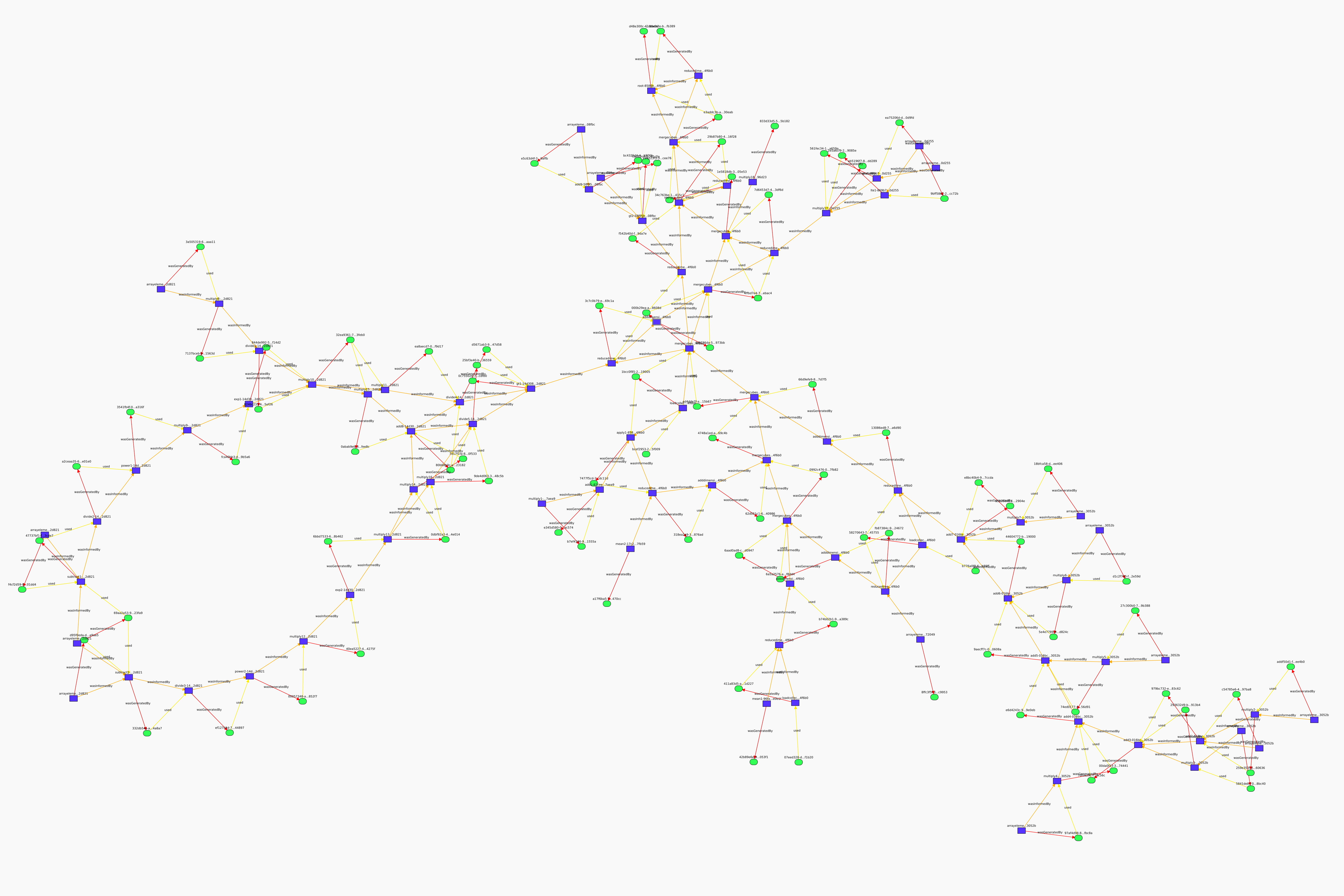}
    \captionsetup{justification=centering}
    \caption{openeo-flood-mapper representation through the yProvExplorer\\
    \scriptsize\protect\url{https://explorer.yprov.disi.unitn.it/?file=http\%3A\%2F\%2Fyprov.disi.unitn.it\%3A3000\%2Fapi\%2Fv0\%2Fdocuments\%2Fyprov4wfs}}
    \label{fig:widefigure}
\end{figure*}

As previously mentioned, openEO offers two distinct modes of use. On one hand, the local processing requires additional setup on the user's machine, including the installation of the openEO Python client and its dependencies. In this mode, the workflow execution is handled entirely by local resources. On the other hand, the remote processing leverages cloud-based infrastructure to execute openEO workflows, offloading the computational tasks from the local environment.

To run the use case on the local machine the following packages have been installed: \texttt{openeo-python-client}; \texttt{openeo-pg-parser-networkx>=2023.5.1}; \texttt{openeo-processes-dask>=2023.7.1}. 

The \textit{openeo-flood-mapper} was run both locally and remotely to process satellite-based flood detection workflows. The yProv4WFs extension for openEO enables full provenance tracking in both cases, allowing detailed inspection of each step of the data flow.

The gathered information is then stored in a JSON file following the W3C PROV standards and it is successively used to generate the corresponding provenance graph. Figure \ref{fig:mapping_funtions} illustrates how the function \texttt{water\_backscatter} in the \textit{openeo-flood-mapper} is mapped to an openEO process graph and subsequently how each sub-function can be plotted in the provenance graph. 

The four different colours, pink, green, orange and red, respectively identify \texttt{plia\_dc} which is the dataset used for the experiment and in the provenance graph it is shown how data are uploaded for the experiment through the invocation of the load function, \texttt{apply} which applies the global linear model, using slope and intercept, to the incidence angle band to estimate the expected water backscatter, \texttt{reduce\_dimension} which reduces the result over time using a mean reducer, and \texttt{add\_dimension} which reintroduces a 'bands' dimension and assigns the label 'wbsc' to the output as a new band for the water backscatter. 

To deeply understand Figure \ref{fig:mapping_funtions}, each operational node in the process graph could have a sub-process graph, which is not represented in the figure for space reasons. Furthermore, each operational node corresponds to an activity in the provenance graph and activities are represented using blue rectangles. Yellow ovals, instead, represent data in input or in output to the different activities, and can be recognized on the  base of the type of connections shown on the directed edges. White rectangles contains extra information about the activity or the entity to which are linked (i.e. execution time and status of an activity or type and dimensions for the entity).This integrated visualization demonstrates how openEO workflows defined in Python code can be systematically parsed, executed, and documented using the yProv4WFs library.

The complete execution of the \textit{openeo-flood-mapper} is shown in Figure \ref{fig:widefigure}. For a better visualization the yProvExplorer service has been exploited, allowing to appreciate the complexity of the use case.

\section{Conclusion and future works}
In this work, we demonstrated how data provenance can be implemented at the workflow level within the openEO platform, supporting both local and remote processing. By integrating the yProv4WFs data model with the parsing stage of openEO workflows, we were able to extract provenance information that contributes to key goals such as transparency, trustworthiness, and repeatability. The Flood Mapper use case was employed as a test scenario to illustrate the extraction of the provenance graph, and the results have been presented in this paper.

The implementation was primarily carried out within the \textit{openeo-pg-parser-networkx} package, which provides an infrastructure-agnostic solution for workflow-level provenance tracking. We have validated that the yProv4WFs data model, developed in alignment with the W3C PROV standards, effectively captures all relevant aspects of provenance in openEO workflows.

Future work will focus on extending this framework by integrating yProv4WFs with additional Earth Observation (EO) platforms to further enhance provenance capabilities across diverse workflow environments.

\section*{Acknowledgment}
This work was funded by the EU HE interTwin project (GA 101058386) and the EU HE Climateurope2 project (GA 101056933). Moreover, this work was partially funded under the NRRP, Mission 4 Component 2 Investment 1.4, by the European Union – NextGenerationEU (proj. nr. CN\_00000013).

\end{document}